\documentclass[twocolumn,showpacs,preprintnumbers,amsmath,amssymb]{revtex4}
 \usepackage{graphicx}
 
\begin{document}
 
\title{Spatially heterogenous dynamics in dense, driven granular flows}
\author{Allison Ferguson, Bulbul Chakraborty}
\affiliation{Martin Fisher School of Physics, Brandeis University,
Mailstop 057, Waltham, MA 02454-9110, USA}
\date{\today}

\begin{abstract}
Interest in the dynamical arrest leading to a fluid $\rightarrow$ solid transition in thermal and
athermal systems has led to questions about the nature of these transitions. These jamming transitions may be dependent 
on the influence of extended structures on the dynamics of the system. Here we show results from a simple driven, dissipative, 
non-equilibrium system which exhibits dynamical heterogeneities similar to those observed in a supercooled liquid which is a 
system in thermal equilibrium. Observations of the time $\tau_R(r)$ required for a particular particle to move a distance $r$ reveal 
the existence of large-scale correlated dynamical regions with characteristic timescales chosen from a broad distribution. The mean 
squared displacement of ensembles of particles with varying characteristic $\tau_R(r)$ reveals an intriguing spatially heterogenous mobility. 
This suggests that a unified framework for jamming will have to be based on the connection between the nature of these heterogeneities 
and the effective dynamics.
\end{abstract}


\maketitle

\section{Introduction} Materials as diverse as molecular liquids, foams and granular matter experience a transition from a fluid-like state to
a solid-like state characterized only by a sudden arrest of their dynamics.  It has been proposed that the phenomena
associated with this dynamical arrest can be explained within the unified framework of a jamming phase diagram~\cite{liu98}.  
Unlike thermal phase transitions, jamming does not seem to be characterized by singularities in any purely static,
structural property~\cite{jaeger96,kadanoff99}. Numerous experiments and simulations indicate the appearance of
large-scale dynamical heterogeneities and an associated growing dynamical length scale. In colloids near the glass
transition, for example, fast-moving particles were observed to be spatially correlated with a characteristic cluster
size that increased as the glass transition was approached~\cite{weeks00}. In supercooled liquids, spatial
inhomogeneities can be identified via a time-dependent four-point (two-time, two-space) correlation
function~\cite{glotzer00}. Recent experiments on molecular liquids and colloids have identified a growing dynamical
length scale near the glass transition via new multipoint dynamic susceptibilities~\cite{berthier05}. In
granular materials, spatial structures have been directly observed in experiments on
flowing systems~\cite{bonamy02, miller96}. More recently, measurements of velocity correlations in the surface layer of
particles flowing down an inclined plane have yielded a length scale which appears to grow as the flow is
arrested~\cite{pouliquen04}. While detection of such dynamical heterogeneities in numerical studies has been more
difficult, extended chains of particles experiencing high frequency collisions are visible in event-driven simulations
of gravity driven granular flow~\cite{ferguson04}.  These structures experience a slow relaxation from collisional
stress at intermediate time scales in a manner analogous to temporal relaxations observed in glassy systems at low
temperatures~\cite{ferguson06}. There is also some evidence for the existence of large-scale structure in images of the
contact force network in simulations of chute flow~\cite{silbert05}. Several questions naturally arise: (1) Are
dynamical heterogeneities a necessary precursor to dynamical arrest of the jamming kind? (2) Do these heterogeneities
and their mesoscopic scale lead to any kind of universal dynamical behavior irrespective of significant differences in
microscopic dynamics? (3) Do static structures such as force chains observed in jammed granular packings emerge out of
dynamical heterogeneities? 

In the present work we address these questions via a simplified model which focuses on the essential effects of driving
and dissipation in a flowing granular system. Simulations have been performed of a two-dimensional gravity driven system
of frictionless bidisperse hard disks in a hopper geometry. The disks undergo instantaneous, inelastic binary collisions and
propagate under gravity in between collisions; a driven, dissipative system that is very different from supercooled molecular
liquids. Previous work~\cite{ferguson04} has shown that this minimal model still reproduces observable results from
related experimental systems~\cite{longhi02}.  To investigate the relevant time and length scales governing the
dynamics of the granular system, an interesting analysis for quantifying dynamical heterogeneity in non-equilibrium
systems~\cite{hurley95} independent of the microscopic interactions between particles is employed.  Specifically, 
measurements of the time $\tau_R(r)$ required for a particular particle to move a specified distance $r$, provide an
appealing physical picture of large-scale correlated regions with characteristic timescales chosen from a broad
distribution.  Through a coarsegraining scheme first proposed in Ref.~\cite{hurley95} which exploits the fact that the
visible heterogeneity is maximized at a particular distance $r_c$, both the distance $r_c$ (which can be associated with
average cage size) and the spatial extent of the correlated regions $\xi^*$ are extracted in a completely
threshold-independent way. $r_c$ can then be used to explore the effect of complex collective behaviour on dynamic
properties such as the average mean squared displacement $\langle (\Delta r)^2
\rangle$.  

\section{Methods} 
The grain dynamics used in the simulations are similar to Ref.~\cite{denniston99}. Specifically, (i) at each
interparticle collision, momentum is conserved but the collisions are inelastic so the relative normal velocity is
reduced by the coefficient of restitution $\mu$; (ii) to allow the side walls to absorb some vertical momentum we impose
the condition that collisions with the walls are inelastic with a coefficient of restitution $\mu_{wall}$ in the
direction tangential to the wall and (iii) since we wish to observe the system over many events, particles exiting the
system at the bottom must be replaced at the top to create uniform, sustained flow.  Note that collisions are
instantaneous and in between collisions the particles are driven by gravity. To avoid the phenomena of inelastic
collapse the coefficients of restitution $\mu$ and $\mu_{wall}$ are velocity dependent; if the relative normal velocity
between particles or between a given particle and the wall is less than some cutoff $v_{cut}$ then the collision is
presumed to be elastic~\cite{bennaim99}.  The flow velocity $v_f$ is controlled by adjusting the width of the hopper
opening. We also introduce a probability of reflection $p$ at the bottom which reduces the time needed to reach
steady-state flow. Typically, our simulations were done on bidisperse systems (diameter ratio 1:1.2) of 1000 disks, with
$\mu$ = 0.8, $\mu_{wall}$ = 0.5, $v_{cut}$ = $1\times10^{-3}$ and $p$ = 0.5. The simulation was run for a total time of
$1\times10^3$ in simulation time units (smaller particle diameter $d_s$, smaller particle mass $m_s$ and gravitational
constant $g$ are all set to 1) with the initial time interval of $5\times10^2$ discarded before recording data to ensure
the system has reached steady state.  During the total time interval of $500$ over which we are evaluating the data, a
given particle will pass through the hopper 5-10 times depending on the flow velocity.

\section{Relaxation times and cage-breaking fluctuations}

\begin{figure}
\begin{center}
\includegraphics[width = 0.95\columnwidth] {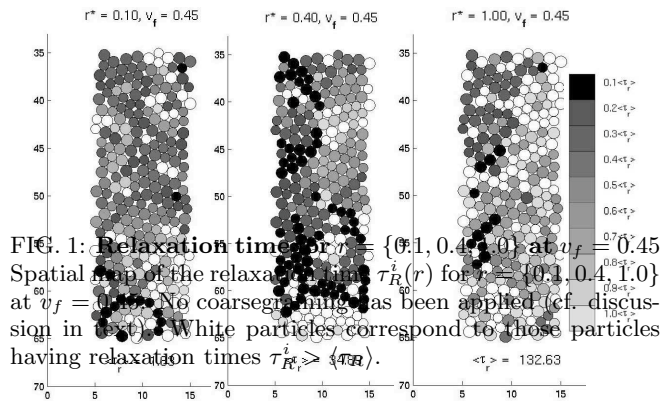}
\caption{\textbf{Relaxation time for $r = \{0.1, 0.4, 1.0\}$ at $v_f = 0.45$} Spatial map of the relaxation
time $\tau_R^i(r)$ for $r = \{0.1, 0.4, 1.0\}$ at $v_f = 0.45$.  No coarsegraining has been applied (cf. discussion in
text). White particles correspond to those particles having relaxation times $\tau_R^i > \langle \tau_R
\rangle$.}
\label{fig:taur}
\end{center}
\end{figure}

Subdiffusive behavior and caging have become hallmarks of the jamming transition, but defining a quantitative measure 
for these processes which is sensitive to the presence of dynamical heterogeneity has been difficult. The following 
observation suggests a possible procedure for defining a characteristic time for a particular particle which may
detect the presence of any spatial heterogeneity in the dynamics: if the relaxation time
$\tau_R^i(r)$ of a given particle $i$ is defined as the time required for particle $i$ to travel a distance $r$, then
any spatial correlation in $\tau_R^i(r)$ (given initial positions of the particles) will depend very strongly on the
chosen value of $r$.  If $r$ is small, then it is not possible to distinguish between those particles which are simply
rattling back and forth within some confined distance and those which may have some enhanced motility.  On the other
hand, if $r$ is too large, then a given particle $i$ may have been in regions where the local average relaxation time
has taken on many values, both fast and slow.  In both cases, a spatial map of the relaxation time reveals no clear
positional correlation in regions of $\tau_R^i(r)$ beyond some small length scale associated with random clumping
(Fig.~\ref{fig:taur}a and c, respectively). 

However, at some intermediate value of $r$, it should be possible to distinguish between localized motion and extended
translational motion prior to a ``washing-out'' of any spatial correlations between particles with similar relaxation
times (compare Fig.~\ref{fig:taur}b, where $r = 0.45$ to Fig.~\ref{fig:taur}a and c, and note the clearer distinction
between regions with different $\tau_R^i(r)$). Presumably there is some optimal value $r_c$ at which the visible
heterogeneity in $\tau_R(r)$ is maximized.  The primary aim of the analysis proposed in Ref.~\cite{hurley95} is to
determine this value of $r_c$ and thence the distribution of ``critical'' relaxation times $P(\tau_R^c)$ where $\tau_R^c
= \tau_R(r_c)$. Particle behaviour such as mean squared displacement can then be analyzed in the context of $\tau_R^c$
for each individual grain and whether or not the particle is then ``fast'' or ``slow''.

\begin{figure}
\begin{center}
\includegraphics[width = 0.95\columnwidth] {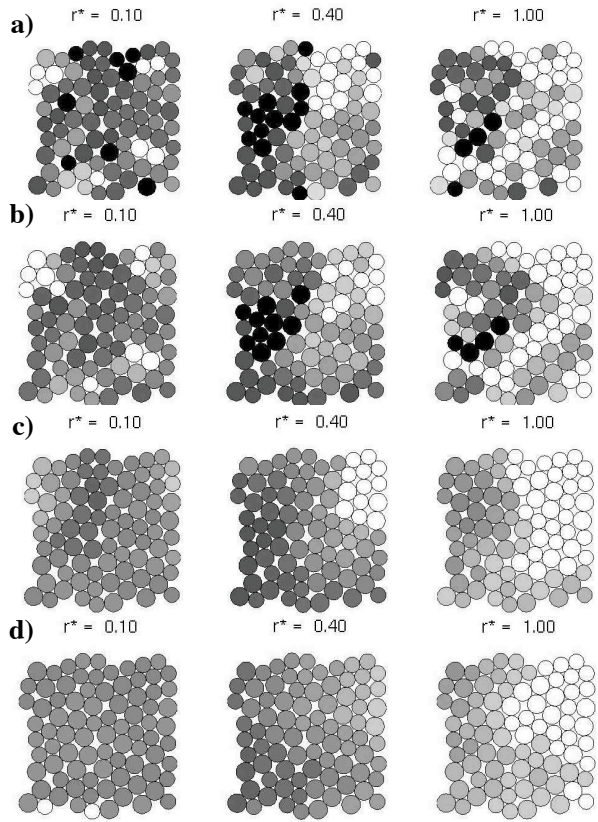}
\caption{\textbf{Effect of coarsegraining on relaxation time map for $r = \{0.1, 0.4, 1.0\}$}Spatial map of the
relaxation time $\tau_R^i(r, l)$ for $r = \{0.1, 0.4, 1.0\}$ at $v_f = 0.45$ for (a) no coarsegraining ($l = 1$), (b) $l
= 2$, (c) $l = 5$ and (d) $l = 10$.  Colour scale is as for Fig.~\ref{fig:taur}.}
\label{fig:taur_cg}
\end{center}
\end{figure}

The method for determining the critical relaxation length $r_c$ is as follows: Define the coarsegrained relaxation time
$\tau_R^i(r, l)$ for a given particle $i$ as the mean local relaxation time in a box of size $l \times l$ centered on
particle $i$. As the size of the coarse-graining region is varied from $l = 1$ (essentially, the single particle
relaxation time) to the size of the analysis region $l = L$, the spatial variation in $\tau_R^i(r, l)$ should decrease
(the result of the coarsegraining procedure on the spatial map is depicted for one snapshot of the system in
Fig.~\ref{fig:taur_cg} for coarsegraining lengths of $l = 2, 5, 10$). Thus, the distribution $P(\tau_R(r, l)$ should become
increasingly narrower, having zero width at $l = L$ where all of the $\tau_R^i(r, L) = \frac{1}{N} \sum_i^N \tau_R^i(r,
1) = \overline{\tau_R}$.  It is clear from the picture that the
heterogenities persist up to larger $l$ for $r \sim 0.4$. If the second cumulant $m_2(r, l)$ of $P(\tau_R(r, l))$
is calculated as:
\begin{equation}
m_2(r, l) = \frac{\langle (\tau_R^i(r, l) - \overline{\tau_R})^2 \rangle}{\langle (\tau_R^i(r, 1) -
\overline{\tau_R})^2 \rangle}
\end{equation}
then $m_2(r, l)$ should be a monotonically decreasing function of $l$, where the length scale associated with the decay
of $m_2(r, l)$ should vary with $r$ and be maximal at $r_c$.

It is worth noting at this point that the original measurement performed in Ref.~\cite{hurley95} was for a 2D periodic
system of soft particles in thermal equilibrium interacting via the repulsive potential $V(r) = \epsilon(\sigma/r)^{-12}$. 
While the validity of this analysis should not depend on the type of microscopic interaction between particles, two
additional considerations had to be taken into account before performing this calculation on the flowing granular
system. Firstly, motion due to the net flow of the system had to be subtracted off prior to calculating $\tau_R^i(r, l)$. 
While more complicated schemes involving a spatially varying flow field could be devised to more accurately compensate
for this effect, for the analysis here the velocity over the region of interest $v_f(t)$ was assumed to be constant in
space at the snapshot at time $t$ (\textit{i.e.} so a particle's displacement $\Delta \mathbf{r}(t)$ in a given time
interval $\Delta t$ is calculated as $\Delta \mathbf{r}(t) = \mathbf{r}(t) - \mathbf{r}(t-\Delta t) - v_f(t-\Delta
t)\Delta t$).  This first approximation seemed to be sufficient to extract $r_c$. Secondly, because relaxation times
were only calculated for particles in the constant flow region, not all particles in the system will be in this region
at a given time, nor will a given particle necessarily remain in the region of interest until it reaches the specified
value of $r$.  So the calculation only considers a given particle trajectory until it departs the analysis region.

\begin{figure}
\begin{center}
\includegraphics[width = 0.95\columnwidth] {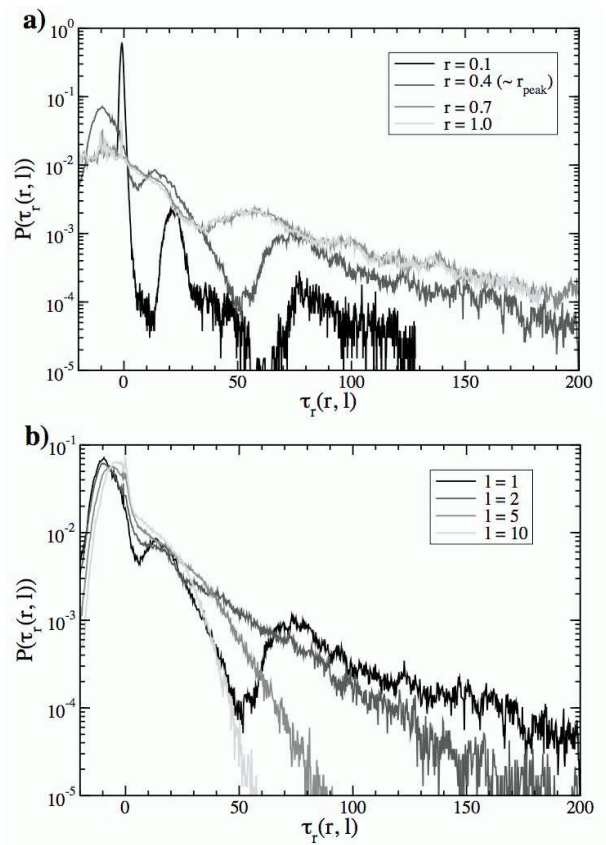}
\caption{\textbf{Relaxation time distributions $P(\tau_R(r, l))$ for $v_f = 1.13$.}Relaxation time distributions $P(\tau_R(r,
l))$ for (a) $r = 0.4$ ($\sim r_c$ for this flow velocity) and (b) $l = 1$ for $r = \{0.1, 0.4, 0.7, 1.0\}$.  In all
cases the distributions are for $v_f = 1.13$ and the average relaxation time $\overline{\tau_R}$ has been subtracted
off. Note that the average collision time for this flow velocity is $\tau \sim 1\times 10^{-3}$ and thus typical
relaxation times are $\gg \tau$.}
\label{fig:trdist}
\end{center}
\end{figure}

Relaxation time distributions $P(\tau_R(r, l))$ for $r = 0.4$ and $l = \{1, 2, 5, 10\}$ are shown in
Fig.~\ref{fig:trdist}a, with a comparison of the single-particle (\textit{i.e.}, $l = 1$) relaxation time for several
values of $r$ in Fig.~\ref{fig:trdist}b. 
The effect of increasing the coarsegraining region for a given value of $r$ is to smooth out the distribution for 
$\tau_R(r, l) > \overline{\tau_R}$, and the width of the distribution decreases as expected for increasing $l$. 
Note that the single particle distribution for $l=1$ can be quite broad for large values of r. 
This is a reasonable result in the context of the calculation of $r_c$; the individual distributions for a particular 
 value of $r \neq r_c$ can be broad initially, but unless there is any significant spatial correlation in relaxation
times the spatial structure of  $\tau_R(r, l)$ will vanish rapidly with increasing $l$ compared to a value of $r$ with
nontrivial spatial correlations. Thus the utility of $\tau_R(r, l)$ lies in its ability to emphasize spatial
correlations between fast and slow particles (\textit{i.e.} if there are particles of widely different mobilities in the
system but they were randomly distributed in space then no interesting correlation length is visible).

\begin{figure}
\begin{center}
\includegraphics[width = 0.95\columnwidth] {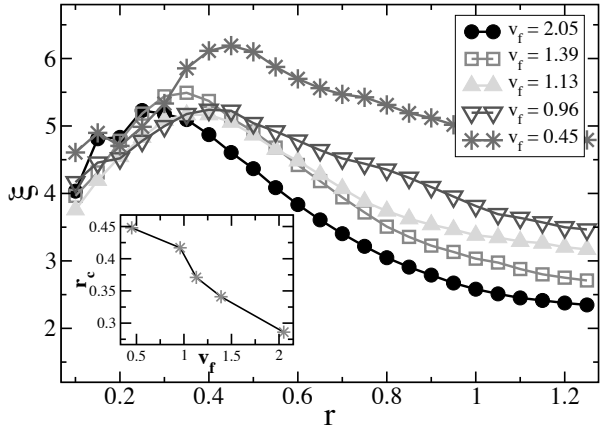}
\caption{\textbf{$\xi(r)$ for varying $v_f$.}Decay length scale $\xi(r) = \int_1^L m_2(l, r) dl$ as a function of $r$
for varying $v_f$.  Inset shows peak location $r_c$ vs. $v_f$.}
\label{fig:xi}
\end{center}
\end{figure}

The length scale associated with the decay of $m_2(l, r)$ can be calculated from $\xi(r) = \int_1^L m_2(l, r) dl$. As
can be clearly seen in Fig.~\ref{fig:xi}, there is a distinct peak in $\xi(r)$ from which two separate and interesting
length scales can be extracted .  One is the location in $r$ of the peak; this quantity is $r_c$ and it is the length
scale over which the dynamical heterogeneity in relaxation times is maximized as described above.  An additional
interpretation can be made of $r_c$ as the critical fluctuation associated with cage-breaking in supercooled fluids
(or equivalently, the typical cage size). This was proposed in Ref.~\cite{hurley95} and confirmed experimentally for
slow, quasistatic granular shear flows in Ref.~\cite{marty05}.  Referring to the inset to Fig.~\ref{fig:xi}, this
quantity is decreasing monotonically with flow velocity, although the exact functional form is difficult to determine.
It is also important to note that because $r_c$ can be related back to a critical relaxation time, changes in $r_c$
with $v_f$ will correspond to changes in the distribution of $\tau_R^c$.

The second length scale which may be determined from $\xi(r)$ is the peak height $\xi^*$, which represents the average
size of the heterogeneity.  From Fig.~\ref{fig:xi}, it is difficult to determine if  $\xi^*$ is varying with $v_f$, or
varying very slowly compared to the changes in $r_c$.  This is in contrast to the results of Ref.~\cite{hurley95} where
an increase in density produced an increase in $\xi^*$ while $r_c$ remained fixed.  Thus the crucial length scale which
is growing as the flow velocity decreases in our system appears to be $r_c$. There is some indication of a crossover
between a low-velocity regime where $\xi$ may be changing more rapidly than $r_c$ to a high-velocity regime where
changes in $r_c$ dominate as described above, but the current data is not sufficient to quantitatively distinguish
between these two possibilities. Additionally, the effect of the finite size of the system on the dynamically correlated regions 
remains an interesting unresolved issue. While the correlation lengths measured over the range of flow velocities measured 
here are still sufficiently smaller than the system dimension, certainly at the slowest flow velocity observed the size of these 
regions begins to approach the system size (at least in the direction perpendicular to the flow).  In order to fully observe the 
behaviour of this length scale as the flow velocity is decreased further, it will be necessary to use larger systems.

\section{Spatially heterogenous mobility} 
\begin{figure}
\begin{center}
\includegraphics[width = \columnwidth] {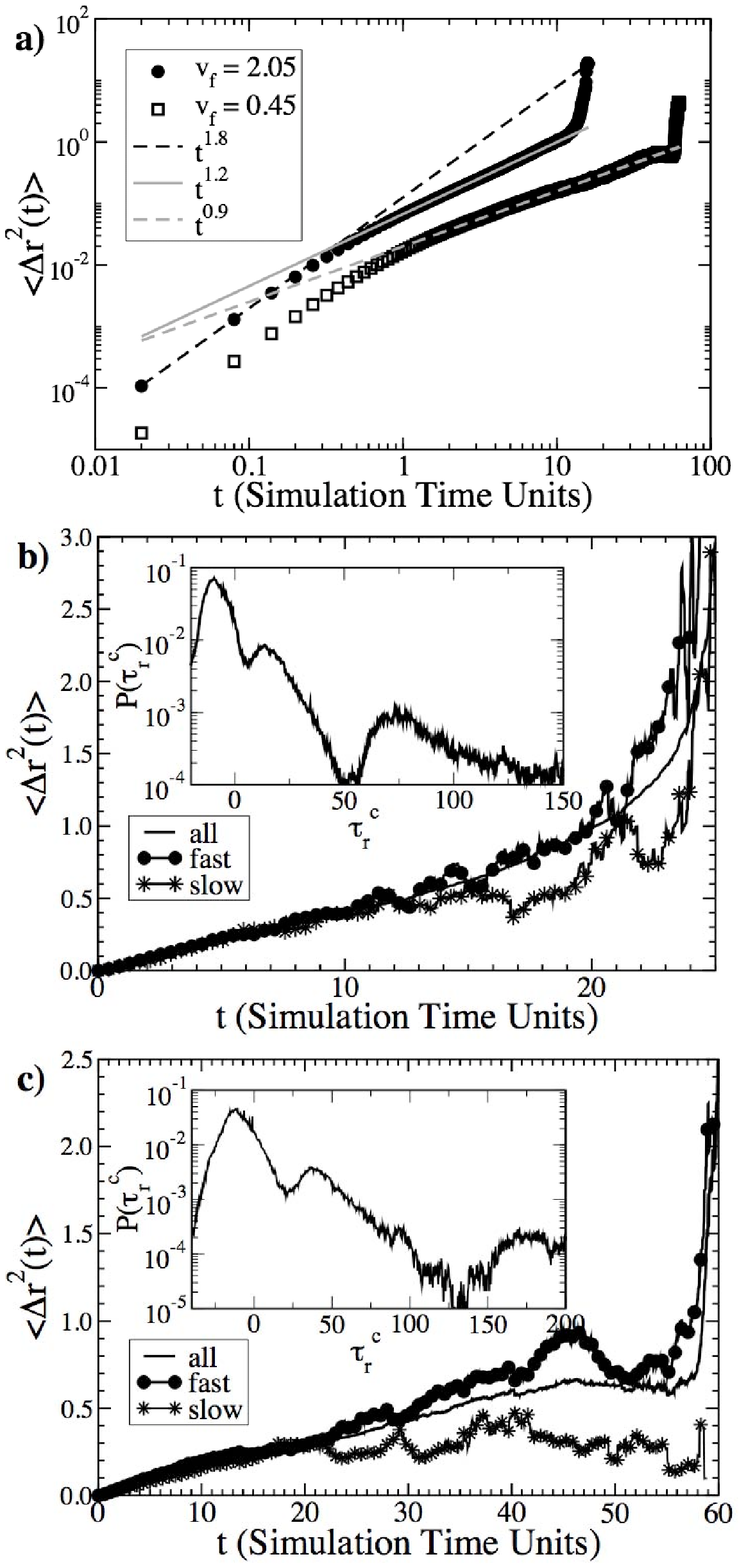}
\caption{\textbf{Mean squared displacement $\langle\Delta r^2\rangle$ for varying $v_f$.} (a) Mean squared
displacement (MSD) $\langle \Delta r^2\rangle$ ($\langle \rangle$ indicates average over all particles in a snapshot at
time $t_0$, then average over all snapshots $t_0$) for $v_f = 2.05$ and $v_f = 0.45$. (b)-(c) Comparison of ``slow''
particles ($\tau_R^c \sim t_{cross}$) and ``fast''particles ($\tau_R^c \sim 1/3 - 1.2 t_{cross}$) to the average over all
particles ({\it cf.} discussion in text) for (b) $v_f = 1.13$ and (c) $v_f = 0.45$.}
\label{fig:msd}
\end{center}
\end{figure}

With the critical fluctuation $r_c$ determined, the critical relaxation time
$\tau_R^c$ for each particle can be defined (Fig.~\ref{fig:taur}b shows a snapshot of the $v_f = 0.45$ system at $r =
r_c$).  As in Ref.~\cite{hurley95}, different particles can now be grouped into subsets according to their value of
$\tau_R^c$ and the contribution of these subsets to dynamical measures (in this case, the mean squared displacement
(MSD) $\langle\Delta r^2\rangle$) can be evaluated. 
Plots of $\langle\Delta r^2\rangle$ vs. time are shown in Fig.~\ref{fig:msd} for varying flow rates (the distributions
of critical relaxation times for a given flow rate are shown in the insets to Fig.~\ref{fig:msd}).  In all cases, the
MSD averaged over all particles in the system in the short-time regime is approximately ballistic ($\langle\Delta
r^2\rangle \sim t^{1.8}$). The long time power-law behaviour is characterized by a flow velocity dependent exponent
($\langle\Delta r^2\rangle \sim t^{\gamma(v_f)}$, where $0.9 < \gamma(v_f) < 1.2$ in the range of flow velocities
studied) which is nearly diffusive.  Fig.~\ref{fig:msd}a shows the power-law fits to the mean squared displacement
curves for two different flow velocities.  At times comparable to the average time it takes a particle to cross the
analysis region $t_{cross} = l/v_f$, the particle falls out of the dense constant flow velocity region and accelerates
under gravity, leading to the sharp increase in the mean squared displacement at $t \sim t_{cross}$.

The MSD curves for ``slow'' ($\tau_R^c \sim t_{cross}$) and ``fast'' ($\tau_R^c \sim (1/3 - 1/2)~t_{cross}$) particles
reveal a deeper complexity not visible in the mean squared displacement averaged over the system as a whole.  As $v_f
\rightarrow 0$, these two groups of particles show increasing separation from the average MSD, with the fast particles
experiencing greater displacement than the average in a given time interval, and the slow particles exhibiting a plateau
(Fig.~\ref{fig:msd}b-c).  For the slowest flow velocity measured, where the plateau is best defined, the plateau occurs
at $\langle\Delta r^2\rangle \sim r_c^2$, corresponding to the experimental result of Ref.~\cite{marty05} where $r_c$
has been directly associated with the cage size. Thus the overall transport in this system is governed by the dynamical
heterogeneity: fast and slow regions are evident as depicted pictorially in Fig.~\ref{fig:taur} with particles in the
slow region experiencing caging.  As the flow velocity decreases, and the length scale $r_c$ increases, the effect of
these mesoscopic regions becomes more pronounced, with the slow particles at $v_f = 0.45$ remaining caged for the
duration of their time to flow through the system (recall that for all of these displacement measurements motion due to
the average flow has been subtracted off).  This picture of rich collective dynamics occuring even in the absence of any
evident density fluctuations was also suggested by Menon and Durian in experiments on dense gravity-driven granular flow
using diffusing wave spectroscopy (DWS)~\cite{menon97}. This is also in analogy to the MSD for the Lennard-Jones system
of Ref.~\cite{hurley95}, where slow particles (defined as the particles in the slowest $40\%$ of $P(\tau_R^c)$) were
seen to be caged over some time $\tau_{mix}$ during which the fast particles experienced diffusion with a diffusion
constant twice that of the assembly of disks as a whole.

\section{Conclusions} In summary, an interesting analysis for quantifying dynamical heterogeneity in non-equilibrium
systems~\cite{hurley95} independent of the microscopic interactions between particles has proven to be a useful tool for
further investigation into the relevant time and length scales governing the dynamics of the gravity-driven granular
system.  Measurements of relaxation time $\tau_R(r)$, defined as the time required for a particular particle to move a
specified distance $r$, provide an appealing physical picture of large-scale correlated regions with characteristic
timescales chosen from a broad distribution of relaxation times $P(\tau_R)$.  Through a clever coarsegraining scheme
which exploits the fact that the visible heterogeneity is maximized at a particular distance $r_c$, both the distance
$r_c$ (which can be associated with average cage size) and the spatial extent of the correlated regions $\xi^*$ were
extracted in a completely threshold-independent way.  While $\xi^*$ was shown to be $\sim 5-6 d$ independent of flow
velocity, the cage size $r_c$ grows as the flow velocity decreases.  $r_c$ then specifically defines the critical
distribution of relaxation times $P(\tau_R^c)$ which can be used to explore the effect of complex collective behaviour
on dynamic properties such as the average mean squared displacement $\langle (\Delta r)^2 \rangle$. The MSD in turn
reveals a nontrivial spatial dependence in the mobility, indicating that while the total mean squared displacement at
long times may seem approximately diffusive, there is underlying complexity in the form of ``fast'' and ``slow'' regions
in the sample.  This result is in agreement with several experimental studies on dense, driven granular
materials~\cite{menon97, marty05} and other jamming systems~\cite{weeks00}. Additionally, there is a qualitative
resemblance between the spatial maps of relaxation time near the critical value $\tau_R^c$ (\textit{c.f.}
Fig.~\ref{fig:taur}b) and recent measures of dynamic propensity and local Debye-Waller factors in 2D simulations of
glass-formers~\cite{cooper06}.\\
One as yet unresolved puzzle is the relationship between the frequent-collision/high-stress chains identified in
previous studies~\cite{ferguson04, ferguson06} and the mesoscopic clumps of fast and slow particles characterized by a
particular value of $\tau_R^c$. From observations of the spatial map of collision frequency and critical relaxation time
at a given snapshot of the system (not shown) it is difficult to ascertain if there is any correlation between the
chains and regions of varying $\tau_R^c$, indeed, there can still be strong heterogeneity in the critical relaxation
time in the absence of any chains.  One measureable connection is that $\langle \tau_R^c \rangle \sim 1/v_f$, which is
the same scaling with flow velocity that was measured in the stress autocorrelations in Ref.~\cite{ferguson06} and
associated with the lifetime of the chains. Thus $\tau_R v_f =$ constant could be an important clue to the dynamical
principles leading to the heterogeneities, and may provide some additional insight into the reasons for why the
frequently-colliding/high-stress chains form in these dense granular systems.

\acknowledgments AF and BC acknowledge support from NSF through grants No. DMR 0207106 and DMR 0549762, and AF acknowledges support
from the Natural Sciences and Engineering Research Council, Canada.

\end{document}